# Low-frequency phase-signal measurement with high-frequency squeezing

Zehui Zhai(翟泽辉)[1,2,*], Jiangrui Gao(郜江瑞)[1,2,3]

[1]State Key Laboratory of Quantum Optics and Quantum Optics Devices, Shanxi University, Taiyuan, 030006, People's Republic of China

[2]College of Physics and Electronics Engineering, Shanxi University, Taiyuan, 030006, People's Republic of China

[3]Institute of Opto-Electronics, Shanxi University, Taiyuan 030006, People's Republic of China

**Abstract**

We calculate the utility of high-frequency squeezed-state enhanced two-frequency interferometry for low-frequency phase measurement. To use the high-frequency sidebands of the squeezed light, a two-frequency intense laser is used in the interferometry instead of a single-frequency laser as usual. We find that the readout signal can be contaminated by the high-frequency phase vibration, but this is easy to check and avoid. A proof-of-principle experiment is in the reach of modern quantum optics technology.

**Introduction**

The concept of radiation field squeezing [1] has attracted much interest since its application to gravitational wave detection was proposed [2]. This sort of radiation field was first generated with a four-wave mixing process by Slusher et al [3] and improved with an optical parametric oscillator (OPO) by Wu et al [4]. It has been useful in various quantum-enhanced measurement schemes, such as squeezed-light magnetometry [5], displacement measurement [6], spectroscopy [7], polarization measurement [8], phase measurement [9], and so on. Squeezed states of light are also vital ingredients for continuous-variable quantum communication and quantum computation [10]. In some applications, such as gravitational wave detection, it is appealing to measure low-frequency signals with a squeezed light state. Low-frequency squeezing has attracted much interest in recent years as terrestrial gravitational wave detectors are approaching their shot noise limit, which can be overcome with squeezing at audio frequencies. Great progress of generating this kind of squeezing has been made in recent years [11–17]. It's found that coherently controlling the phases of the experimental set-up while not to introduce extra noise is one of the keys to generate low-frequency squeezing [13-17]. In 2007, an unprecedented experiment of generating squeezed vacuum states with a noise power 6.5 dB below vacuum noise within the entire detection bandwidth of ground-based GW-detectors (10 Hz - 10 kHz) was demonstrated by using a sophisticated control scheme [17]. An alternative way to conquer this difficulty is to use a two-frequency laser and broadband squeezing at higher frequency, which has been primarily used in many quantum optics laboratories, to enhance the signal-to-noise ratio (SNR) of an interferometer for lower-frequency phase-signal measurement. In 1987, Yurke et al [18] proposed a squeezed-state enhanced two-frequency interferometer to perform sub-shot-noise measurement of low-frequency signals by reading the photocurrent at frequency $2\nu_s$, well away from the low-frequency technical noise. In this study, we read low-frequency signals directly and calculate the SNR. We find that the readout signal can be contaminated by high-frequency phase vibrations, but this is easy to check. This technique can be straightforwardly extended to the other squeezing-enhanced measurement schemes mentioned above.

---

[*] Electronic address: zhzehui@sxu.edu.cn

**Theoretical model**

We consider a squeezing-enhanced Mach-Zehnder interferometer as in Ref. [9] shown in Fig. 1. The main idea is that by transferring the frequency of the intense laser to the frequencies of the entangled sidebands of the squeezed field, the low-frequency signal originating from the sidebands close to the intense laser frequencies can be detected with enhanced SNR for the entangled squeezing sidebands. The frequency relations of the usual scheme and our scheme are shown in fig. 2. Suppose the two-frequency laser and squeezed state have annihilation operators $\hat{A} = \hat{a}_+ e^{-i(\omega_0+\Omega)t} + \hat{a}_- e^{-i(\omega_0-\Omega)t}$ and $\hat{B} = \hat{b} e^{-i\omega_0 t}$ respectively, where $\omega_0$ is the optical circular frequency and $\Omega$ is the circular frequency, at which squeezing occurs. With the mean-field approximation, the annihilation operators $\hat{a}_+$, $\hat{a}_-$ and $\hat{b}$ can be expressed as the sum of its mean field and fluctuation, i.e. $\hat{a}_i = \alpha_i + \delta\hat{a}_i$ $(i=+,-)$, and $\hat{b} = \delta\hat{b}$. Here, squeezing is supposed to be vacuum-squeezing, so that the corresponding mean field is zero. The mean amplitudes $\alpha_+ = \alpha_- = \alpha$ are real numbers by choosing a proper phase reference and are supposed to be equal. The average photon numbers in unit measurement time is $N = 2\alpha^2 = P/\hbar\omega_0$, where $P$ is the optical power of the intense two-frequency laser.

With the relative phase $\varphi$ at the first 50% beam splitter BS1 and relative phase $\pi/2+\theta(t)$ at the second 50% beam splitter BS2 in the Mach-Zehnder interferometer, the annihilation operators of the various modes in fig. 1 are related to each other via

$$\hat{C} = \frac{1}{\sqrt{2}}\left(\hat{A}+\hat{B}e^{i\varphi}\right), \tag{1}$$

$$\hat{D} = \frac{1}{\sqrt{2}}\left(\hat{A}-\hat{B}e^{i\varphi}\right), \tag{2}$$

and

$$\hat{E} = \frac{1}{\sqrt{2}}\left(\hat{C}+\hat{D}e^{i(\pi/2+\theta(t))}\right)$$
$$= \frac{1}{2}\left[\left(1+e^{i(\pi/2+\theta(t))}\right)\hat{A}+\left(1-e^{i(\pi/2+\theta(t))}\right)\hat{B}e^{i\varphi}\right], \tag{3}$$

$$\hat{F} = \frac{1}{\sqrt{2}}\left(\hat{C}-\hat{D}e^{i(\pi/2+\theta(t))}\right)$$
$$= \frac{1}{2}\left[\left(1-e^{i(\pi/2+\theta(t))}\right)\hat{A}+\left(1+e^{i(\pi/2+\theta(t))}\right)\hat{B}e^{i\varphi}\right], \tag{4}$$

where $\theta(t)$ is a sum of the low-frequency cosine signal $\theta_{w_1}\cos w_1 t$ to be measured at frequency $w_1$ and some spurious signals which could be coupled to the photocurrent. $\theta(t)=\sum_s \theta_{w_s}\cos w_s t$, where $\theta_{w_s}$ is the signal amplitude at frequency $w_s$. The frequency spacing is much larger than the measurement resolution bandwidth $\Delta\omega$. The quantum efficiency of photodiodes is supposed to be unity, so that the subtracted output photocurrent deduced from eqs.(1)–(4) is

$$\hat{i}(t) = \hat{F}^\dagger\hat{F}-\hat{E}^\dagger\hat{E}$$
$$= N\sum_s \theta_{w_s}\cos w_s t\left(1+\cos 2\Omega t\right)+\sqrt{2N}\delta\hat{X}_b^{\varphi+\pi/2}(t)\cos\Omega t$$
$$= s(t)+n(t), \tag{5}$$

where the quadrature fluctuations are defined as $\delta\hat{X}_b^\varphi = \delta\hat{b}e^{i\varphi}+\delta\hat{b}^\dagger e^{-i\varphi}$, and the product terms of quadrature fluctuations and the terms multiplying $\theta(t)$ and quadrature fluctuations are omitted for $\theta \ll 1$ and $\delta X_i^\varphi \ll \alpha$ $(i=+,-,b)$. We take the photocurrent duration as $T$ $(1/T \ll w_s)$, which is the reciprocal of measurement resolution bandwidth $\Delta\omega/2\pi$. The photocurrent $i(t)$ is separated into a signal part $s(t)$ and noise part $n(t)$, and the power spectral densities are calculated separately. By using the Wiener-Khinchine theorem, the power spectral density of the first term of eq.(5), i.e. the signal term, is

$$P_s(w) = \int_{-\infty}^{\infty} d\tau e^{-iw\tau} R(\tau) = \frac{1}{T}\int_{-\infty}^{\infty} d\tau e^{-iw\tau} \int_{-T/2}^{T/2} dt s(t)s(t-\tau)$$

$$= \frac{\pi N^2}{2}\sum_s \left(2\theta_{w_s}^2 + \theta_{w_s}\theta_{w_s-2\Omega} + \theta_{w_s}\theta_{w_s+2\Omega}\right)\left(\delta(w+w_s)+\delta(w-w_s)\right)$$

$$+\frac{\pi N^2}{4}\sum_s \left(\theta_{w_s}^2 + \theta_{w_s}\theta_{w_s+4\Omega} + 2\theta_{w_s}\theta_{w_s+2\Omega}\right)\left(\delta(w+w_s+2\Omega)+\delta(w-w_s-2\Omega)\right)$$

$$+\frac{\pi N^2}{4}\sum_s \left(\theta_{w_s}^2 + \theta_{w_s}\theta_{w_s-4\Omega} + 2\theta_{w_s}\theta_{w_s-2\Omega}\right)\left(\delta(w-w_s+2\Omega)+\delta(w+w_s-2\Omega)\right), \quad (6)$$

where we use the approximation that $\sin\omega T/(\omega T) \approx 1$ when $\omega=0$ and $\sin\omega T/(\omega T) \approx 0$ when $\omega \neq 0$. Function $R(\tau) = \frac{1}{T}\int_{-T/2}^{T/2} dt s(t)s(t-\tau)$ is the autocorrelation function of signal $s(t)$. Function $\delta(w-w_1)$ is the Dirac delta function defined as $\int_{-\infty}^{\infty} dt e^{-iwt} = 2\pi\delta(w)$, and satisfies $\int_{w_1-\Delta\omega/2}^{w_1+\Delta\omega/2} dw\delta(w-w_1)=1$. Also, we use $\theta_{w_s} = \theta_{-w_s}$. The noise power spectral density of the second term on the other side of eq.(5), i.e. the noise term, is

$$P_n(w) = \left\langle \frac{1}{T}\int_{-\infty}^{\infty} d\tau e^{-iw\tau} \int_{-T/2}^{T/2} dt n(t)n(t-\tau) \right\rangle_{en}$$

$$= \left\langle \left|\mathcal{F}_T\{n(t)\}\right|^2 \right\rangle_{en} / T$$

$$= \frac{N}{2}\left\langle \int_{-T/2}^{T/2} dt \delta X_b^{\varphi+\pi/2}(t)\left(e^{-i(w-\Omega)t} + e^{-i(w+\Omega)t}\right)\int_{-T/2}^{T/2} dt' \delta X_b^{\varphi+\pi/2}(t')\left(e^{i(w-\Omega)t'} + e^{i(w+\Omega)t'}\right)\right\rangle_{en} / T$$

$$= \frac{N}{2}\left[V_b^{\varphi+\pi/2}(w+\Omega)+V_b^{\varphi+\pi/2}(w-\Omega)\right], \tag{7}$$

where $\langle...\rangle_{en}$ refers to the ensemble average, and $\mathcal{F}_T\{n(t)\}=n_T(w)=\int_{-T/2}^{T/2}dt n(t)e^{-iwt}$ is the Fourier transform of function $n(t)$. $V_b^{\varphi+\pi/2}(w+\Omega)$, defined as

$$V_b^{\varphi+\pi/2}(w+\Omega)=\left\langle\left|\delta X_b^{\varphi+\pi/2}(w+\Omega)\right|^2\right\rangle_{en}/T=\left\langle\left|\int_{-T/2}^{T/2}dt\delta X_b^{\varphi+\pi/2}(t)e^{-i(w+\Omega)t}\right|^2\right\rangle_{en}/T \tag{8}$$

is the quadrature variance of the squeezed state at frequency $w+\Omega$. To evaluate the SNR at frequency $w_1$, we integrate the signal spectral density (eq. (6)) and noise spectral density (eq. (7)) in the frequency interval $[w_1-\Delta\omega/2,\ w_1+\Delta\omega/2]$, and take the ratio as the SNR:

$$\begin{aligned}P_s(w_1)&=\int_{w_1-\Delta\omega/2}^{w_1+\Delta\omega/2}\frac{dw}{2\pi}P_s(w)\\&=N^2\left(\theta_{w_1}+\theta_{w_1+2\Omega}/2+\theta_{w_1-2\Omega}/2\right)^2,\end{aligned}\tag{9}$$

and

$$\begin{aligned}P_n(w_1)&=\int_{w_1-\Delta\omega/2}^{w_1+\Delta\omega/2}\frac{dw}{2\pi}P_n(w)\\&=\frac{N\Delta\omega}{4\pi}\left[V_b^{\varphi+\pi/2}(w_1+\Omega)+V_b^{\varphi+\pi/2}(w_1-\Omega)\right].\end{aligned}\tag{10}$$

Therefore

$$SNR=\frac{P_s(w_1)}{P_n(w_1)}=\frac{2TN\left(\theta_{w_1}+\theta_{w_1+2\Omega}/2+\theta_{w_1-2\Omega}/2\right)^2}{V_b^{\varphi+\pi/2}(w_1+\Omega)+V_b^{\varphi+\pi/2}(w_1-\Omega)}. \tag{11}$$

The integration of equ.(10) takes the quadrature variances to be uniform in the integral frequency interval. Expressions (9) and (11) show that the phase vibrations at frequencies $w_1-2\Omega$ and $w_1+2\Omega$ will contaminate the output photocurrent at frequency $w_1$. Therefore, the photocurrent signal at $w_1$ may not faithfully reflect the phase vibration at $w_1$. However, when the phase vibrations at $w_1-2\Omega$ and $w_1+2\Omega$ do not exist, the SNR of eq.(11) can be simplified to

$$SNR = \frac{2NT\theta_{w_1}^2}{V_b^{\varphi+\pi/2}(w_1+\Omega)+V_b^{\varphi+\pi/2}(w_1-\Omega)}. \tag{12}$$

In this case, the photocurrent at frequency $w_1$ does reflect the phase vibration at $w_1$ and has sub-shot noise resolution if the quadrature variables of the squeezed state at the angle $\varphi+\pi/2$ are squeezed at frequencies $w_1-\Omega$ and $w_1+\Omega$. Moreover, it is experimentally easy to check with some accuracy whether the phase signals at frequencies $w_1-2\Omega$ and $w_1+2\Omega$ exist by using the intense laser at central frequency $\omega_0$ instead of the two-frequency laser. The effects of some specific signals at frequencies $w_1-2\Omega$ and $w_1+2\Omega$ can be avoided by changing the frequency $\Omega$ in the squeezing bands. Therefore, by using a two-frequency laser, the measurement of phase signals at low frequency $w_1$ can be improved with squeezing at high frequencies $w_1-\Omega$ and $w_1+\Omega$ when the squeezing angle is properly aligned.

The physics of this scheme can be explained as follows. Instead of using a single-frequency intense laser at frequency $\omega_0$ in the interferometry, we use an intense laser with the carrier frequencies $\omega_0-\Omega$ and $\omega_0+\Omega$, with which the entangled upper and lower sidebands of vacuum squeezing interfere. Therefore, the low-frequency signal, which generates sidebands near the carrier fields, can be detected with sub-shot noise resolution, thanks to the entangled sidebands of squeezing. The terms with phase signals at frequencies $w_1-2\Omega$ and $w_1+2\Omega$ coupled to the SNR at frequency $w_1$ in eq. (11) originate from the beating between one of the carriers and the sidebands at $w_1-2\Omega$ and $w_1+2\Omega$ of the other carrier. It's interesting to note that in eq.(11) the signal appears at frequency $w_1$ and

$w_1 \pm 2\Omega$ while the noise at frequency $w_1 \pm \Omega$. This difference can be attributed to the fact that, in Mach-Zehnder interferometer, signal sidebands are generated around carrier fields while noise sidebands come from the injected squeezing.

**Conclusion**

We have calculated the utility of high-frequency squeezed-state enhanced two-frequency interferometry for low-frequency phase measurement. By means of a two-frequency laser interferometer, the higher-frequency sidebands of the squeezed state can be used to enhance the lower-frequency phase signal measurement. The subsequent photocurrent signal can be contaminated by higher-frequency phase vibrations, but this can be easily checked and avoided. A proof-of-principle experiment is in the reach of modern quantum optics technology and is in progress in our laboratory. Moreover, this scheme is also useful for many other squeezing-enhanced measurement schemes, and also provides a method to generate low-frequency squeezing with high-frequency squeezing.

**Acknowledgments**

The authors thank the referee's critical comments and suggestion. Zehui Zhai thanks Dabo Guo for helpful discussion. This work was supported in part by the National Science Foundation of China under Grant No. 11174189, No.60708010 and No.60978008, by the National Basic Research Program of China under Grant No.2010CB923102, and by the Open project of State Key Laboratory of Quantum Optics and Quantum Optics Devices under Grant No.200902.

Figure Captions:

Fig.1. Squeezed-state enhanced Mach-Zehnder interferometer. OPA: optical parametrical amplifier. BS1 and BS2: 50% beam splitters.

Fig. 2. Frequency relations of two-frequency intense laser, squeezing and signal sidebands for: (a) usual interferometer; (b) our scheme. $w_1$ is the low frequency to be measured. The shadowed areas are entangled sidebands of the squeezed state. Long arrows are carrier frequencies of the intense laser, and the short arrows are signal sidebands.

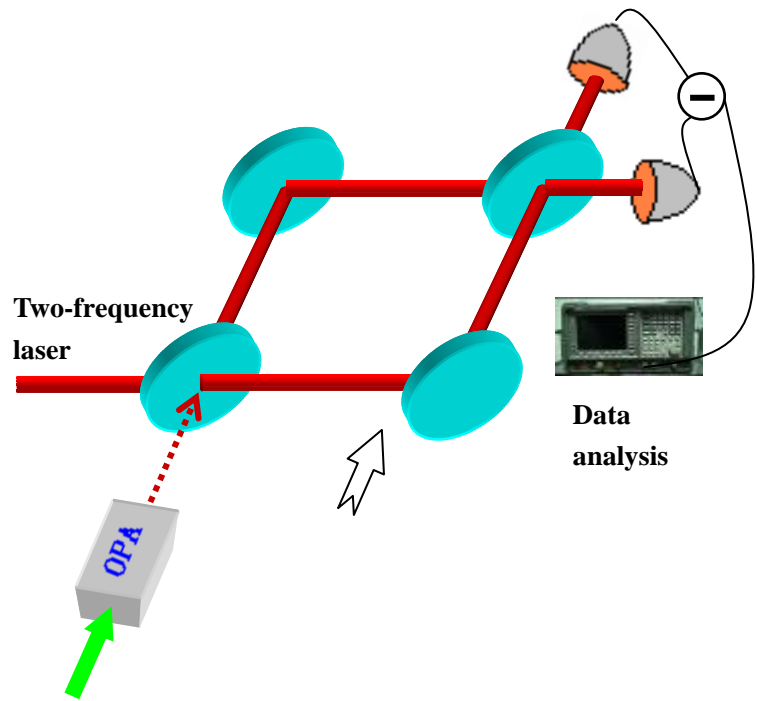

Fig.1. Squeezed-state enhanced Mach-Zehnder interferometer.

OPA: optical parametrical amplifier. BS1 and BS2: 50% beam splitters.

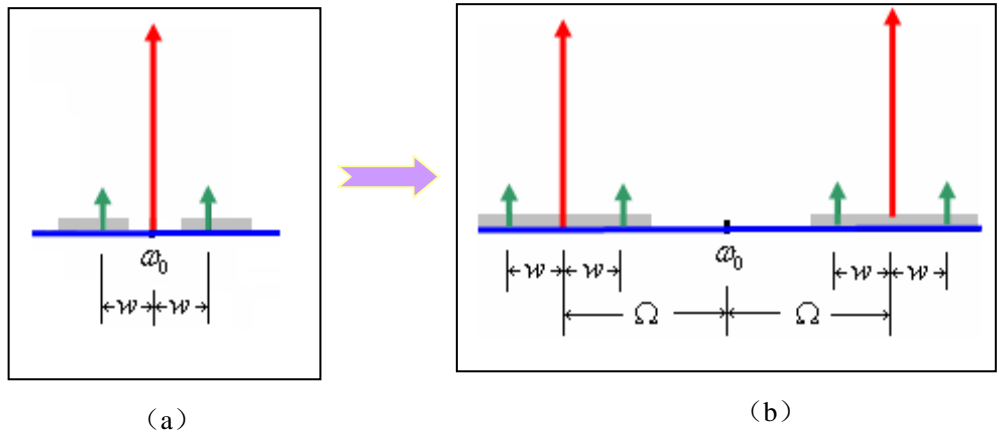

Fig. 2., Frequency relations of two-frequency intense laser, squeezing and signal sidebands for: (a) usual interferometer; (b) our scheme. $w_1$ is the low frequency to be measured. The shadowed areas are entangled sidebands of the squeezed state. Long arrows are carrier frequencies of the intense laser and the short arrows are signal sidebands.